\documentclass[apj]{emulateapj}

\shorttitle{Modeling the Stellar Populations in CMa}
\shortauthors{de Jong et al.}

\begin{document}

\title{Modeling the Stellar Populations in the Canis Major Over-Density:
  the Relation Between the Old and Young Populations}

\author{J. T. A. de Jong, D. J. Butler and H-W. Rix}
\affil{Max-Planck Institut f\"ur Astronomie, K\"onigstuhl 17, 69117 Heidelberg, Germany}
\email{dejong@mpia.de}

\and

\author{A. E. Dolphin}
\affil{Steward Observatory, University of Arizona, 933 N. Cherry Ave,
  Tucson, AZ, 85721, United States}

\and

\author{D. Mart\'inez-Delgado}
\affil{Instituto de Astrof\'isica de Canarias, C/ V\'ia L\'actea,
  E38200, La Laguna, Spain}

\begin{abstract}
We analyze the stellar populations of the Canis Major stellar
over-density, using quantitative color-magnitude diagram (CMD) fitting
techniques. The analysis is based on photometry obtained with the Wide
Field Imager at the 2.2m telescope at La Silla for several fields near
the probable center of the over-density. A modified version of the
MATCH software package was applied to fit the observed CMDs, enabling
us to constrain the properties of the old and young stellar
populations that appear to be present. For the old population we find
[Fe/H]$\sim$-1.0, a distance of $\sim$7.5 kpc and a line-of-sight
depth $\sigma_{los}$ of 1.5$\pm$0.2 kpc and a characteristic age range
of 3-6 Gyrs. However, the spread in ages and the possible presence of a
$\sim$10 Gyr old population cannot be constrained.  The young
main-sequence is found to have an age spread; ages must range from a
few hundred Myr to 2 Gyr. Because of the degeneracy between distance
and metallicity in CMDs the estimates of these parameters are strongly
correlated and two scenarios are consistent with the data: if the
young stars have a similar metallicity to the old stars, they are
equidistant and therefore co-spatial with the old stars; if the young
stars have close to solar metallicity they are more distant ($\sim$9
kpc). The relatively low metallicity of the old main-sequence favors
the interpretation that CMa is the remnant of an accreted dwarf
galaxy. Spectroscopic metallicity measurements are needed to determine
whether the young main-sequence is co-spatial.
\end{abstract}


\section{Introduction}
\label{sec:intro}

A few years ago \cite{martin04a} discovered an apparent over-density of
M-giant stars in the direction of the constellation Canis Major
(CMa), using 2MASS data. They also reported that the structure was
surrounded by a number of globular and open clusters. Combined with
the fact that it is located at low galactic latitudes, just below the
plane of the disk of the Milky Way, led them to conclude that it may
well be the (remnant of a) disrupted dwarf galaxy that could have
spawned the so-called Monoceros stellar stream. That stellar stream
had been discovered some years prior \citep{newberg02,yanny03} and
seems to surround the Galaxy completely at very low galactic
latitudes. Numerical simulations show that such a system can be
explained by an in-plane accretion event \citep[e.g.][]{monmodel}.
Subsequent deep photometric observations
\citep{martinez05,bellazzini04}, kinematic studies
\citep{martin04b,martin05} and analysis of 2MASS data
\citep{bellazzini06} are in accord with this scenario. The optical
photometry also revealed the presence of a young main sequence
population of stars potentially at the same distance. The initial
analysis of a new wide-field survey of the CMa region revealed that
the projected distribution of the young stars is qualitatively similar
to the older population \citep{butler06}.

However, other explanations for the observed stellar over-density have
been offered. \cite{momany04} and \cite{momany06} argued that the
observed stellar over-densities towards CMa reflect the warp and flare
of the outer disk. \cite{carraro05} and \cite{moitinho06} argued that
the intrinsic substructure of the disk such as spiral arms can explain
the observations if they are not in the plane of the disk. In this
picture the young main sequence (YMS) population is related to the
outer Cygnus spiral arm, while the older main sequence (OMS)
feature is caused by disk stars in the inter-arm region between the
Cygnus and Perseus spiral arms. In the latter scenario the OMS and YMS
populations would be located at different distances and not actually
co-spatial.

The generic problem with the interpretation of the observed features
is that since they are lying so close to the plane of the disk it is a
priori not clear if they are intrinsic to the disk or have an outside
origin.

We are performing a large photometric survey of the CMa over-density
using the Wide Field Imager (WFI) on the ESO/MPG 2.2m telescope at La
Silla. Based on a relatively straightforward analysis of the
color-magnitude diagrams (CMDs) we have estimated the line-of-sight
(l.o.s.) depth of the over-density and its spatial extent in
\citep[][B06]{butler06}.  The picture that emerged is that of an old
stellar over-density that is highly elongated in Galactic longitude
with a projected aspect ratio of $\gtrsim$5:1, consistent with recent
2MASS analyses \citep{bellazzini06, rocha06}. The YMS stars show
qualitatively a similar projected distribution but are markedly more
localized, both in galactic longitude and latitude with a maximum near
($l$,$b$)$\sim$(240\degr,-7\degr). For the extent of the
over-density along the l.o.s., $\sigma_{los}$, B06 found upper limits
of 1.8 and 1.5 kpc for the OMS and YMS respectively, assuming an
average distance of 7.5 kpc for both populations.

The main aim of this paper is to extract more information from the
photometry by applying more sophisticated CMD-fitting methods.  In
particular we are interested whether the OMS and YMS over-density
populations seen towards CMa can be co-spatial, based on our optical
photometry.  We also constrain the metallicity of the YMS stars, which
has so far not been constrained. To do this we apply CMD-fitting
techniques to a small selection of fields towards the likely maximum
of the CMa over-density from our large photometric survey.

Fitting of CMDs with model stellar populations enables use of the full
distribution of stars within a CMD. Methods to obtain detailed
information on the star formation and chemical enrichment history of
composite stellar systems have been proven to be very successful
\citep[see e.g.][]{gallart96, tolstoy96, aparicio97, dolphin97,
holtzman99, olsen99, hernandez00, harris01,match}. For the analysis in
this paper a modified version of the software package MATCH
\citep[e.g.][]{dolphin97,match} is used. It uses maximum-likelihood
techniques to find the linear combination of single-component stellar
population models that best fits an observed Hess diagram (stellar
density as function of color and magnitude).

The remainder of the paper is organized as follows: Section
\ref{sec:datameth} describes the data and methods used. In Sections
\ref{sec:scfits} and \ref{sec:distfits} the results of our fits are
presented and their implications are discussed in section
\ref{sec:discussion}. Finally, section \ref{sec:conclusions} presents
our conclusions.

\section{Data and methods}
\label{sec:datameth}

\begin{figure}[t]
\epsscale{1.2}
\plotone{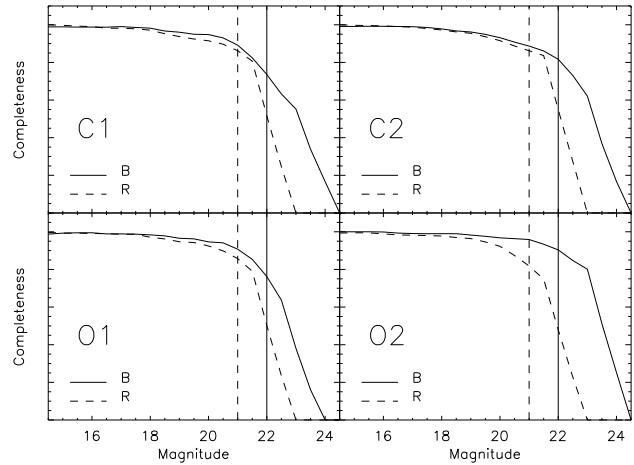}
\caption{Completeness of the stellar photometry for the four fields
  (see section \ref{sec:fields}) used in the current analysis,
  determined from the artificial star tests. Solid lines are for B
  band and dashed lines for R band. The limiting magnitudes we use for
  our CMD fits, B=22 and R=21, are indicated with vertical lines.
}
\label{fig:completeness}
\end{figure}

This analysis is based on deep B,R photometry obtained with the Wide
Field Imager (WFI) at the ESO/MPG 2.2m telescope at La Silla. These
data are part of an extensive survey of the Canis Major region,
presented in B06. Because of the amount of data involved,
the data reduction used for the complete survey as described by
B06 is fully automated. For the analysis in this paper,
accurate calibration of the photometry is critical. We use the same
reduced image data, which are overscan, bias, flat-field and
astrometrically corrected using a pre-reduction pipeline
\citep{schirmer03}, 
but individually re-calibrate the magnitudes for our CMa fields to
ensure the best possible photometric accuracy.

Standard star fields taken on the same night (2004 December 9) as
three of the CMa fields were used to determine the zero-points, color
terms and airmass correction factors. The values found were very close
to the ones provided by ESO. Based on these standard star measurements
we estimate the color calibration accuracy to be 5\%.  Object
detection was performed using SExtractor \cite{sextractor}, after
which PSF-fitting photometry was done with the IRAF\footnote{Image
Reduction and Analysis Facility} task DAOPHOT/ALLSTAR.  Fake star
tests were also done by adding artificial stars to the images using an
empirical PSF measured from stars in the image and subsequently
running the same detection and photometry routines. Six sets of 1000
artificial stars spread over a wide magnitude range (15-24 for B-band,
14-23 for R-band) were used for each image, giving 6000 fake stars per
image. For the analysis presented here we will only go down to
magnitudes of 22 in B and 21 in R. Figure \ref{fig:completeness} shows
that the data are complete to $\sim$80\% complete or better at these
limits. Typical photometric errors at these magnitudes are 0.05 mag
and better at brighter magnitudes.

The CMD fitting software needs artificial star files with large
numbers of stars spread over the whole magnitude and color range
used. B-band and R-band artificial star data are randomly combined
within a color range of -1.0$<$B-R$<$3.0 to create lists of more than
80,000 entries for each field.

\subsection{Fields used and extinction}
\label{sec:fields}

\begin{figure}[t]
\epsscale{1.0}
\plotone{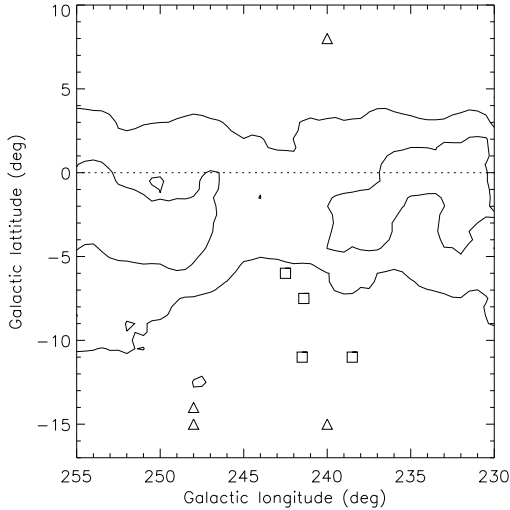}
\caption{Locations of the fields used for the current analysis, where
  squares indicate the CMa fields and triangles the control
  fields. The dotted line indicates the Galactic plane at $b$=0\degr and
  the contours correspond to constant differential extinction values of
  $E(B-V)=1.0$ and 0.4 mag, based on the dust opacity maps by \cite{sfd}. }
\label{fig:locations}
\end{figure}

\begin{figure*}
\epsscale{1.0}
\plotone{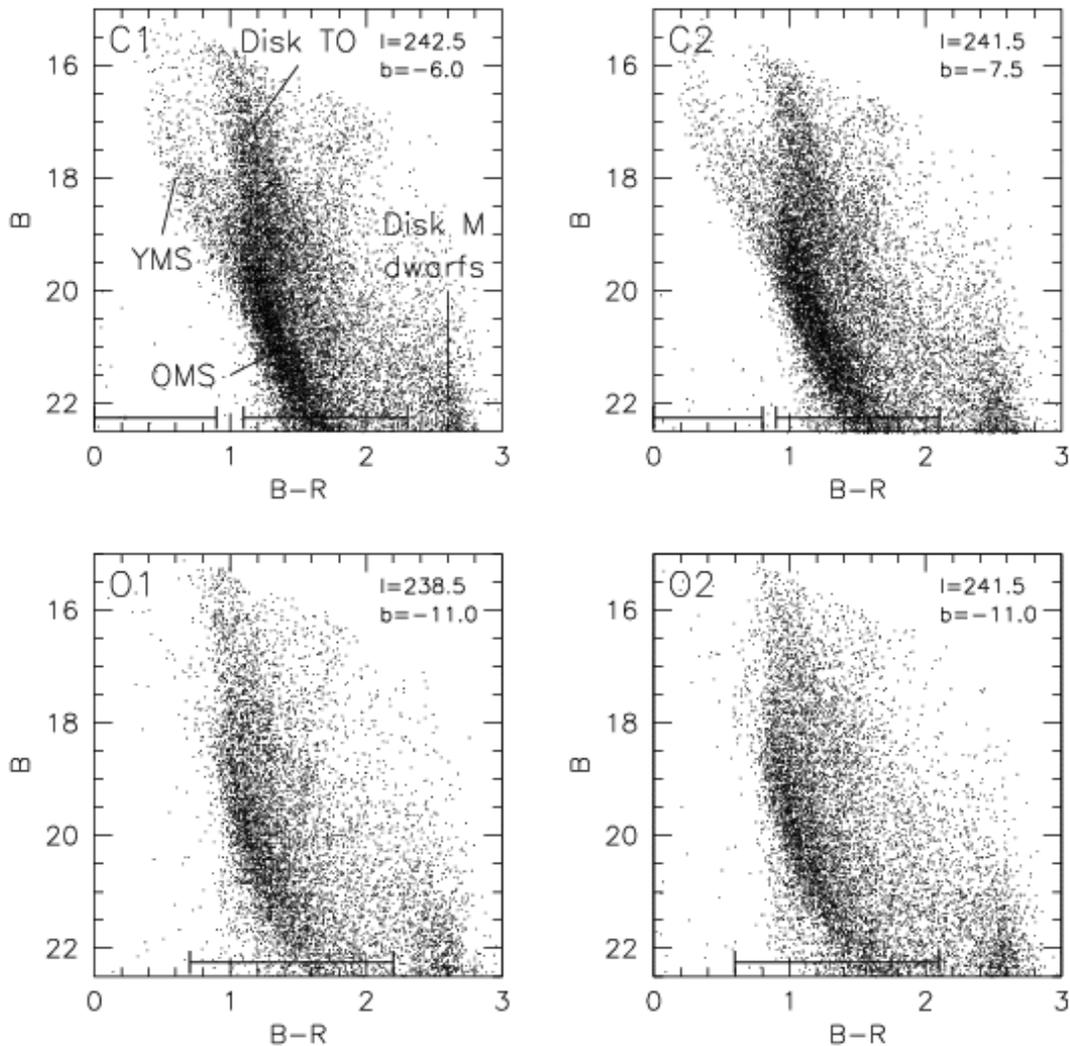}
\caption{Color-magnitude diagrams (CMD) of the CMa fields selected for
  the present analysis. The upper panels show the central fields C1
  (left) and C2 (right) and the lower panels the outer fields O1
  (left) and O2 (right), respectively. Each field is $\sim$0.2
  sq.deg. in area. At the bottom of each CMD the color ranges used
  later for the separate fits to the young and old stars are
  indicated. In the upper left panel the obvious features are
  labeled. A near-vertical swath of predominantly thin and thick disk
  main-sequence turn-off stars is present in all fields, labeled with
  ``Disk TO''. At the bottom right of the CMD, ``Disk M dwarfs''
  indicates the location of a concentration of thin and thick disk M
  dwarfs. The young main sequence (YMS) and old main sequence (OMS) of
  CMa are also labeled.}
\label{fig:cmacmds}
\end{figure*}

\begin{figure*}
\epsscale{1.0}
\plotone{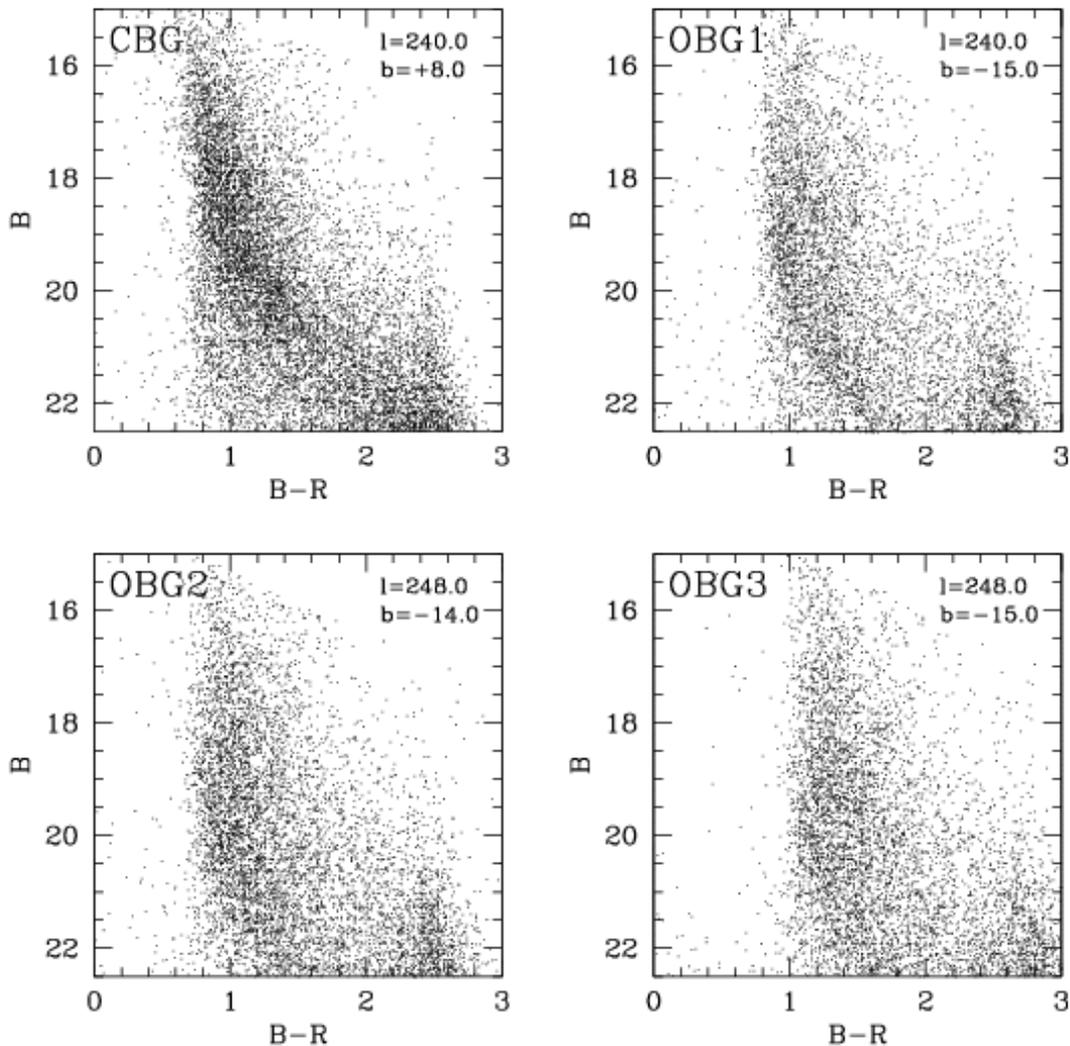}
\caption{Color-magnitude diagrams of the control fields selected for the current analysis. In absence of CMa stars, the halo, thick disk and thin disk features (see Figure \ref{fig:cmacmds}) are clearly seen.}
\label{fig:controlcmds}
\end{figure*}

\begin{deluxetable}{lccc}
\tablecaption{Overview of used fields}
\tablewidth{0pt}
\tablehead{ \colhead{Field} & \colhead{$l$ (\degr)} & \colhead{$b$ (\degr)} & \colhead{$<A_{V,SFD}>$ (mag)}}
\startdata
\multicolumn{4}{c}{CMa Fields}\\
C1 & 242.5 & -6.0 & 1.13 [0.85,1.28]\\
C2 & 241.5 & -7.5 & 0.76 [0.63,1.03]\\
O1 & 238.5 & -11.0 & 0.54 [0.49,0.64]\\
O2 & 241.5 & -11.0 & 0.42 [0.38,0.49]\\
\\
\multicolumn{4}{c}{Control Fields}\\
CBG & 240.0 & 8.0 & 0.40 [0.35,0.42] \\
OBG1 & 240.0 & -15.0 & 0.32 [0.30,0.39] \\
OBG2 & 248.0 & -14.0 & 0.71 [0.61,0.83] \\
OBG3 & 248.0 & -15.0 & 0.61 [0.52,0.82]
\enddata
\tablecomments{C1, C2, O1 and O2 are the CMa fields that are the
  subject of the present analysis. CBG is the control field used for
  fields C1 and C2; OBG1, OBG2 and OBG3 are used to construct control
  field CMDs for O1 and O2. Column four gives the median extinction
  towards each field, with in brackets the maximum and minimum values
  \citep{sfd}.}
\label{tab:fields}
\end{deluxetable}

As shown by B06, our imaging survey traces the CMa
over-density over a large area ($\sim$80\degr$\times$20\degr), but
for this first CMD analysis we limit ourselves to a small number of
fields.  To maximize signal-to-noise (the number of CMa stars vs. the
number of fore- and background stars) and since the YMS is spatially
less extended than the OMS (B06), we chose to use
fields near the presumed center of the over-density at
($l$,$b$)=(240\degr,-7\degr). In order to see whether the
properties of the OMS change with galactic latitude we also want to
study some fields farther away from the plane. To be able to check for
internal errors we choose to use 2 fields near the presumed center of
CMa and 2 fields that are at a latitude of $b\simeq-11$\degr.  Since
the CMa over-density is located at very low galactic latitudes, just
south of the plane of the disk, the number of fore- and background
disk stars is very high. This necessitates the use of
control fields in our analysis, and appropriate fields need to be
selected for this purpose as well.

Because of the location of the CMa over-density close to the Galactic
plane, the extinction in the observed fields is high and increasing
along the l.o.s.. Furthermore, the extinction correction in B06 has
shown that the extinction estimates from \cite{sfd} and
\cite{bonifacio00} in some cases seem to overestimate and in others to
underestimate the actual extinction. This is problematic for our
analysis since an unknown extinction introduces large uncertainties in
the CMD-fitting analysis due to the degeneracies between extinction,
age, metallicity and distance. To minimize these uncertainties we
select fields with relatively low (but certainly still significant)
extinction for our analysis.

Table \ref{tab:fields} lists the coordinates and average extinction
values of the selected fields, where C1 and C2 correspond to the
``central'' fields, i.e. near the CMa center ($b\approx$-7\degr),
and O1 and O2 to the ``outer'' fields ($b=$-11\degr). Also listed
are the fields that will be used to construct control fields. For
fields C1 and C2 a field at similar longitude and similar latitude
north of the disk plane will be used for control field; this field
shows no significant contribution of CMa stars and is listed in Table
\ref{tab:fields} as CBG.  Since fields farther away from the plane
contain much less stars, we combine CMDs from three fields at low
latitude to construct control fields for the fields O1 and O2. These
three control fields (OBG1, OBG2 and OBG3 in Table \ref{tab:fields})
are selected to contain no or very few CMa stars and to have similar
extinction as fields O1 and O2. The locations of the selected fields
are plotted in Figure \ref{fig:locations}, together with contours
indicating the extinction in the Galactic plane.  In Figure
\ref{fig:cmacmds} the CMDs of the CMa target fields are shown, and in
Figure \ref{fig:controlcmds} the CMDs of the control fields.

\subsection{CMD fitting methods}
\label{sec:cmdfitting}

As mentioned in Section \ref{sec:intro}, quantitative and algorithmic
CMD (or Hess diagram) fitting methods have proven themselves and
several software packages have been developed for this aim
\citep{gallart96, tolstoy96, aparicio97, dolphin97, holtzman99,
olsen99, hernandez00, harris01}. The software package we use is MATCH,
which was written and successfully used for the CMD analysis of
globular clusters and dwarf galaxies \citep{match}. The software works
by converting the observed CMD into a Hess diagram and comparing that
with Hess diagrams of model populations. A Hess diagram is a
two-dimensional histogram of stellar density as function of color and
magnitude. Theoretical isochrones \citep[][in the case of
MATCH]{girardi02} are convolved with a model of the photometric
accuracy and completeness, obtained from the artificial star test
data, to create the model Hess diagrams. The use of Hess diagrams
enables a pixel-by-pixel comparison and a maximum-likelihood technique
is used to find the best-fitting linear combination of models. To
account for contamination by fore- and background stars, a control
field CMD can be provided. MATCH will then use the control field Hess
diagram as an extra `model population'. The resulting output will be
the best-fitting linear combination of population models plus the
control field. Previously MATCH was used in a mode where extinction
and distance are fixed and the best-fitting combination of models with
different age and metallicity are found. For the application described
in this paper, we have used MATCH in somewhat different modes, namely
one where distance is a free parameter and one where only simple,
single component models are fit to the data and compared based on
their goodness-of-fit.

MATCH was originally developed for systems where all stars can be
assumed to be at the same distance, i.e. external galaxies. In the
case of the CMa over-density this approximation does not apply, as it
is relatively nearby and has a significant extent along the l.o.s.
\citep[e.g.][]{martinez05,butler06}. Moreover, it is possible that the
OMS and YMS populations are only co-spatial in projection and actually
at different distances \citep{moitinho06}.  Thus, there are two
effects we need to cope with, namely distance spread of populations
and distance offsets between populations.  Accounting for a distance
spread can be done in a straightforward way by manipulating the
photometric error model; adding random offsets with a normal
distribution to the recovered magnitudes of the artificial stars will
smooth the model Hess diagram along the magnitude axis with the same
$\sigma$.  To fit populations at different distances simultaneously we
have implemented a new way in which to run MATCH that keeps the same
number of free parameters, or parameter dimensions, to be explored. In
the standard mode, the distance is fixed, while age and metallicity
are independent variables. Here, the distance becomes a free parameter,
but age and metallicity are being limited via a specified age
metallicity relation (AMR). For any given AMR there is only one
``population parameter'', e.g. age with a certain metallicity linked
to each age bin. Of course, different AMRs can be explored, ultimately
permitting a wide range of age-metallicity relations.  In general,
AMRs should be chosen that are appropriate for the specific
application, based on expected or known metallicity evolution. For
example, for the study of structures in the Galactic thin disk one
would consider a different AMR than for the study of a metal-poor
dwarf galaxy. However, within this new mode of running MATCH the
choice of AMR is very flexible and a flat AMR (no dependence of
metallicity on age) or even an AMR where metallicity decreases with
time is possible. In section \ref{sec:distfits} of this paper we will
use two different AMRs to constrain how the assumed metallicity
influences the distances found through CMD-fitting.

Because of the large contamination of the CMa CMDs with fore- and
background stars, a feature like the red giant branch (RGB) is
observed at very low contrast. The main sequence (MS) is the only
feature that is observed with high contrast and significance, and
MATCH can fit the location of the MS and the MS turn-off (MSTO), as
well as the density of stars along the MS. To constrain the basic
parameters of CMa -- distance, age range, and average metallicity --
we first use MATCH in a single component (SC) fitting mode; in this
mode, template stellar population models are created that consist of a
single component with fixed distance, foreground extinction, age range
and metallicity range\footnote{While we use 'single components' it is
important to note that these are not single stellar populations (SPP)
with only a single age, distance, and chemical composition}. Such a SC
template is fit to the data together with a control field. Since the
template consists of a single component, the only free parameters
during the fit are the absolute scaling of (i.e. the number of stars
in) the template and the control field. For each template (i.e. for
each combination of distance, foreground extinction, age range and
metallicity range), taken by itself, a value of the goodness-of-fit is
obtained. Based on the goodness-of-fit values we can then find the
best-fitting template and thus constrain the overall properties of the
CMa YMS and OMS populations. A paper describing the application of
these new modes of application of MATCH, including tests and
simulations is in preparation (J.T.A. de Jong et al. 2007, in
preparation).

It should be noted that the results of CMD fits depend on the assumed
set of theoretical stellar evolution models. When MATCH was written,
the only set of isochrones with full age and metallicity coverage
available was provided by \cite{girardi02}. Since MATCH only uses one
set of isochrones, this adds a further source of uncertainty to our
fit results.

From the CMDs presented here and in B06, the OMS and YMS seem to be
two distinct populations. In Section \ref{sec:scfits} we first study
the populations separately using SC template fits. To better study the
relation between these two populations, we also fit them
simultaneously using the new capability of MATCH to solve for
distance. Section \ref{sec:distfits} describes the results of these
distance distribution fits to the complete CMa CMDs.

\section{Single component fits to the CMDs}
\label{sec:scfits}

\begin{deluxetable*}{lcccc|cc}
\tablecaption{Single Component Fitting Results}
\tablewidth{0pt} \tablehead{ \colhead{Parameter} & \colhead{C1 OMS} &
  \colhead{C2 OMS} & \colhead{O1 OMS} & \colhead{O2 OMS} & \colhead{C1 YMS} &
  \colhead{C2 YMS}}
\startdata
$B-R$ range (mag) & [1.1,2.3] & [0.9,2.1] & [0.7,2.2] & [0.6,2.1] & [0.0,0.9] & [0.0,0.8]\\
A$_{V,fg}$ (mag) & 1.2$\pm$0.1 & 0.9$\pm$0.2 & 0.7$\pm$0.2 & 0.5$\pm$0.2 & 0.9$\pm$0.1 & 0.6$\pm$0.1 \\
Age$_{ll}$ (log(yr)) & 9.5$\pm$0.1 & 9.4$\pm$0.1 & 9.5$\pm$0.1 & 9.5$\pm$0.1 & 8.4$\pm$0.3 & 8.4$\pm$0.0 \\
Age$_{ul}$ (log(yr)) & 9.7$\pm$0.1 & 9.7$\pm$0.1 & 9.7$\pm$0.1 & 9.7$\pm$0.1 & 9.3$\pm$0.0 & 9.3$\pm$0.0 \\
$[Fe/H]$ (dex) & -1.0$\pm$0.2 & -1.1$\pm$0.3 & -0.6$\pm$0.3 & -0.6$\pm$0.3 & -0.4$\pm$0.2 & -0.3$\pm$0.2 \\
m-M (mag) & 14.35$\pm$0.08 & 14.45$\pm$0.09 & 14.2$\pm$0.2 & 14.3$\pm$0.2 & 14.8$\pm$0.3 & 14.9$\pm$0.2 \\
$\sigma_{los}$ (mag) & 0.44$\pm$0.06 & 0.39$\pm$0.05 & 0.44$\pm$0.06 & 0.39$\pm$0.07 & 0.44$\pm$0.07 & 0.31$\pm$0.07 \\
\enddata
\tablecomments{Age$_{ll}$ and Age$_{ul}$ stand for the lower limit and
  upper limit of the best-fit age range. The $B-R$ range corresponds
  to the color range in the observed Hess-diagram that is used in the
  SC fits to single out either the OMS or the YMS. It differs from
  field to field mainly because of differences in extinction. There
  are no YMS results for fields O1 and O2, as in these fields the YMS
  is absent. The errors given here are the standard deviations in the
  values given by all acceptable fits to the data.}
\label{tab:scresults}
\end{deluxetable*}

\subsection{Fit setup}

The OMS is present in all four CMa fields, but with highest
signal-to-noise (S/N) in the fields near the presumed density maximum
of CMa at ($l$,$b$)$\simeq$(240\degr,-7\degr).  To only fit the
OMS we use color cuts to isolate the regions of the CMDs where the OMS
population is located. The color range used for each field is listed
in Table \ref{tab:scresults} and indicated in Figure
\ref{fig:cmacmds}.  For C1 and C2 we use the CBG control field as
background, where small photometric offsets are applied to CBG to
better match the extinction of each field. Since the exact extinctions
are uncertain this is done ``by eye'', using the locations of the blue
edge of disk MSTO stars and the clump of M dwarfs as a guide.  For O1
and O2 we create background CMDs by applying separate offsets to
control fields OBG1, OBG2 and OBG3 and then combining them.  To test
how sensitive our results are to the exact control field match, tests
were performed where the photometric offsets of the control fields
were varied by 0.05 magnitudes in color and 0.15 magnitudes in
brightness.  This influences the recovered metallicities as presented
later on by at most $\sim$0.1 dex, the foreground extinction by up to
0.1 magnitudes, and the distance modulus also by at most 0.1
magnitudes; the age estimates are not significantly affected.

A range of SC models is generated to be compared with the observed
Hess-diagrams and each has a specific age range, metallicity, distance
modulus, l.o.s. spread, and foreground extinction as detailed below.
In all our models we assume a realistic binary fraction for
main-sequence disk stars of 0.6 \citep{duquennoy91,kroupa93}. 
For fitting the OMS we use 14 metallicity bins varying from
[Fe/H]=-2.05 to -0.1 in steps of 0.15 dex and in all cases there is
uniform metallicity spread of 0.2 dex.  Distance moduli sampled run
from 13.5 to 15.5 mag in steps of 0.1 mag, a range including the
distance measured previously for the OMS
\citet{martin04b,martinez05,bellazzini06} and the distance to the
outer spiral arm, possibly the distance of the YMS.
Extinctions vary from $A_V$=0.5 to 1.4 mag in steps of 0.1 mag for
C1 and C2 and from $A_V$=0.2 to 1.0 mag in steps of 0.1 mag for O1 and
O2. Age bins probed vary both in average age as well as in age spread
and star formation is assumed to be constant during the entire width
of an age bin; we use seven narrow bins with ages in log(years) of 9.4
to 9.5, 9.5 to 9.6, 9.6 to 9.7, 9.7 to 9.8, 9.8 to 9.9, 9.9 to 10.0
and 10.0 to 10.1; seven broader age bins run from 9.5 to 9.7, 9.7 to
9.9, 9.9 to 10.1, 9.5 to 9.8, 9.8 to 10.1, 9.3 to 9.7, and 9.7 to
10.1, respectively. This total of 14 age bins covers ages from $\sim$2
Gyr to $\sim$12.5 Gyr, while the narrowest reflects and age spread of
only 650 Myr and the widest presumes ongoing star formation for 7.5
Gyrs.  To probe the l.o.s. depth of the OMS, each model is also
broadened in magnitude with a Gaussian kernel of a certain
$\sigma_{los}$. We probe values 0.2 to 0.6 mag in steps of 0.05 mag;
at a distance modulus of 14.5, a step of 0.05 magnitudes corresponds
to $\sim$0.18 kpc.

The YMS is only present in the central fields C1 and C2.  For fitting
we use the part of the CMDs blue-ward of $B-R\sim1$ where this is the
predominant population, see Table \ref{tab:scresults} and Figure
\ref{fig:cmacmds} for the exact color ranges. Again we probe the
metallicity range [Fe/H]=-2.05 to -0.1 in steps of 0.15 dex and with a
metallicity spread of 0.2 dex. Distance moduli again range from 13.5
to 15.5 mag with a 0.1 magnitude resolution.  Extinctions vary from
$A_V$=0.5 to 1.4 mag in steps of 0.1 mag.  Twelve age ranges are
probed, namely seven narrow ones from 6.6 to 7.05, 7.05 to 7.5, 7.5 to
7.95, 7.95 to 8.4, 8.4 to 8.85, 8.85 to 9.3, and 9.3 to 9.75, and five
broader ones from 6.6 to 7.5, 7.5 to 8.4, 8.4 to 9.3, 6.6 to 8.4, and
7.5 to 9.3, all in log(years). Again we try different l.o.s. depths in
steps of 0.05 magnitudes. There is little contamination by fore- and
background stars for $B-R<1$ and especially for field C2 the YMS stars
are easily separated from the contaminants. In the C1 CMD the YMS
stars do overlap slightly with young, blue thin disk stars and to take
this into account a control CMD was created based on the CBG control
field.

The quality of each fit is determined using the maximum-likelihood
technique described in \cite{match}. This statistic is meant to
measure how likely it is that the observed distribution of stars would
result from a random drawing from the fitted model. In the current
analysis, we know that our fits will not be perfect since the control
field is not perfect and the SC models we use will not be a perfect
representation of the CMa stellar populations. We therefore need a
relative way of comparing the goodness-of-fit of each model.  Monte
Carlo simulations using random drawings from the models and bootstrap
tests where new CMDs are created by drawing at random individual stars
from the target CMD are performed for this purpose. From the results
the expected 1 $\sigma$ spread in the goodness-of-fit for each field
is determined. We then assume that all model fits that have a
goodness-of-fit value within 1 $\sigma$ of the best fit are
statistically good fits to the data.

\begin{figure*}
\epsscale{1.0}
\plotone{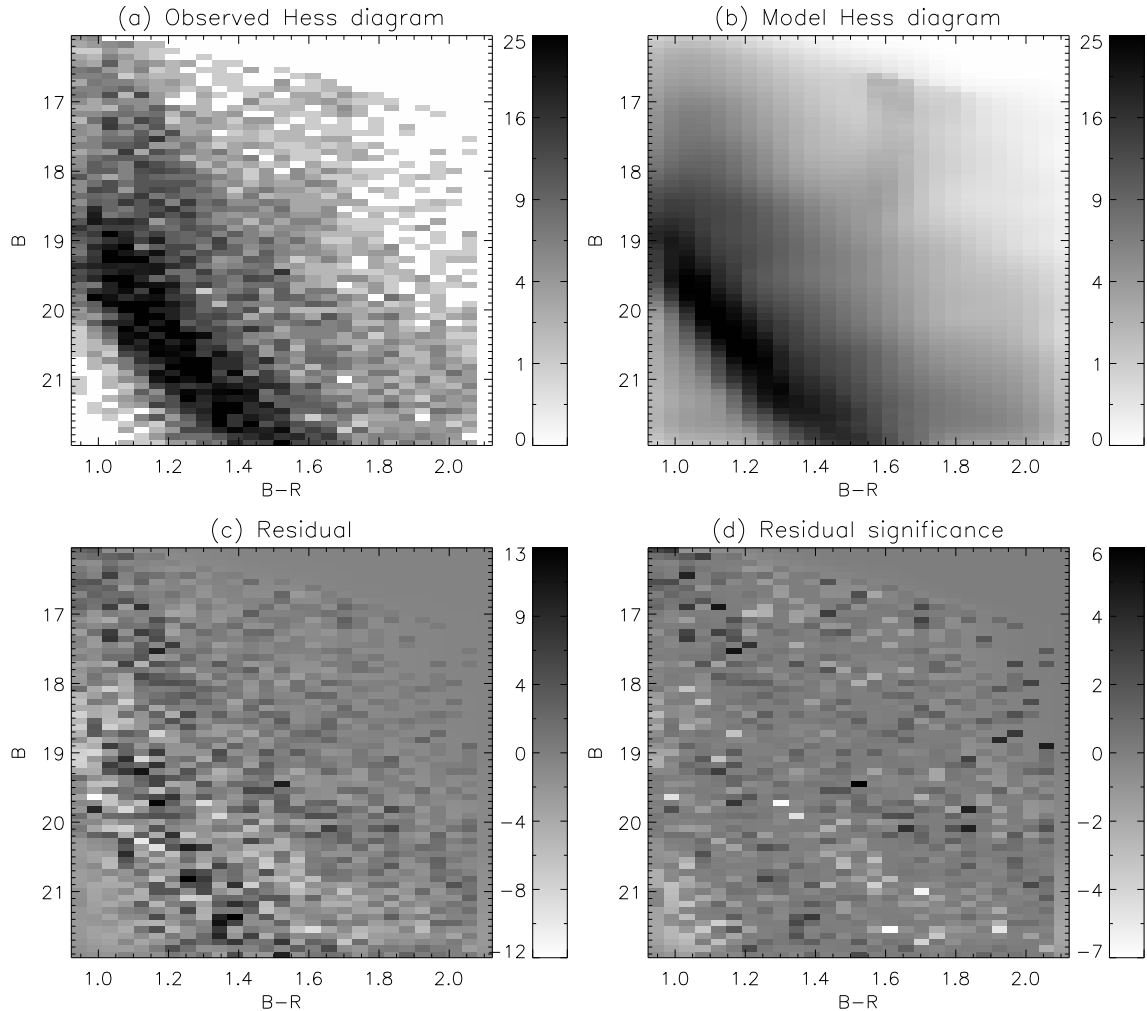}
\caption{
SC fit to the OMS in field C2. (a) Observed Hess diagram; the grey level
gives the number of stars in each color-magnitude box with dark
corresponding to high density. (b) Modeled Hess diagram, consisting of
the background CBG and the best-fitting SC model. (c) Fit residuals,
where light color corresponds to bins where the data is high and dark
to bins where the model is high. (d) Significance of the fit
residuals; this is a function of the discrepancy between data and
model and the total number of stars in the bin. Important is that even
when using a relatively simple SC model there are no significant
systematic residuals.
}
\label{fig:c2_omsfits}
\end{figure*}

\begin{figure*}
\epsscale{1.0}
\plotone{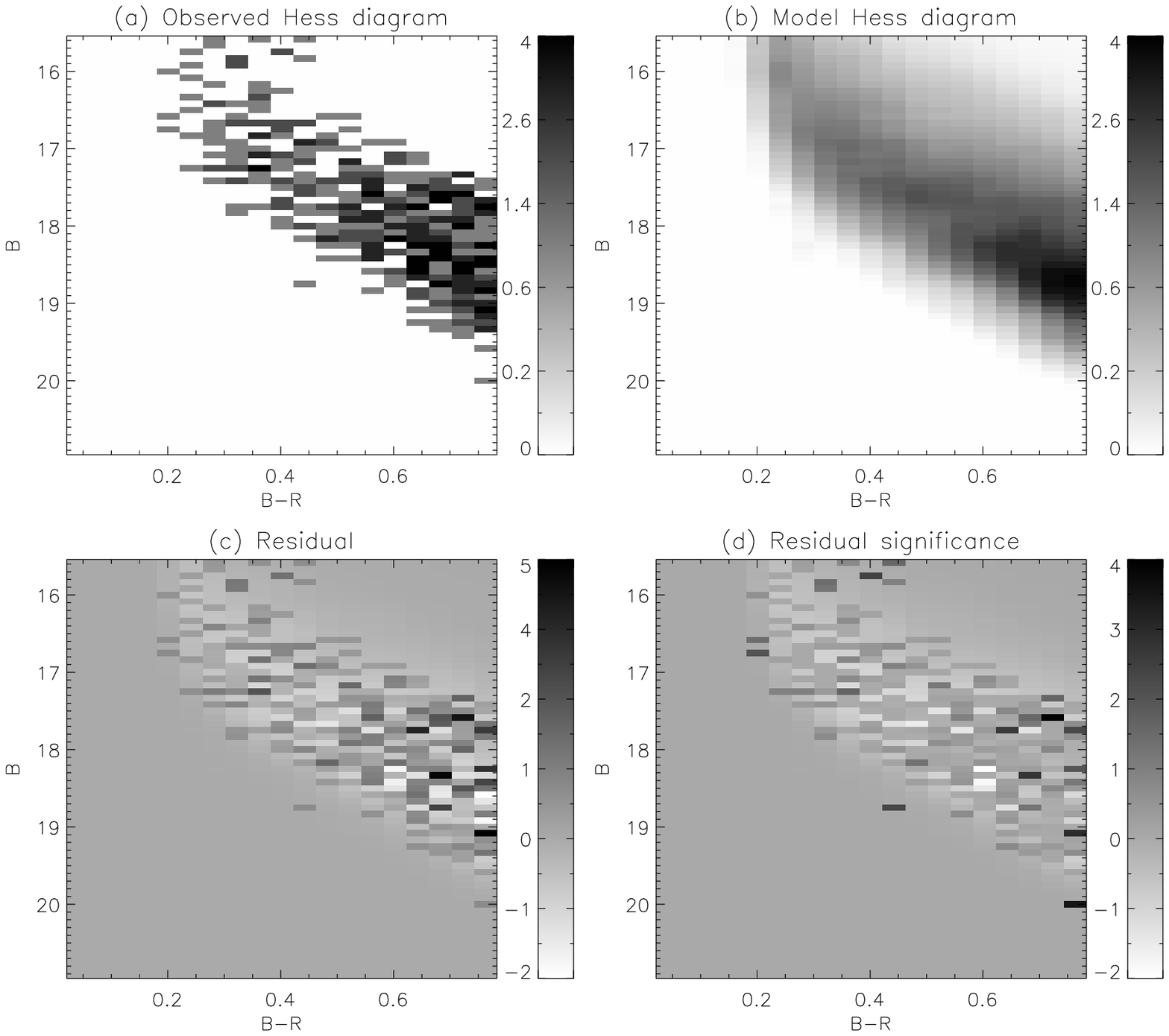}
\caption{
Like Figure \ref{fig:c2_omsfits} but for the best SC fit to the YMS in
field C2.
}
\label{fig:c2_ymsfits}
\end{figure*}

To illustrate the SC fitting procedure, we show in Figure 
\ref{fig:c2_omsfits} the observed Hess diagram, the synthetic
Hess diagram (background + CMa model) of the best fit, the residuals
between the observed and synthetic Hess diagrams and the residual
significance for the OMS in field C2 as an example. It shows
that the observed OMS feature can be modeled well with a relatively
simple SC model. Figure \ref{fig:c2_ymsfits} shows the same for the
YMS in field C2.

\subsection{Results}

Table \ref{tab:scresults} presents the CMa properties we derive for
the OMS and YMS in each field using the above methods. The quoted
values are the unweighted averages of the values of all fits that lie
within 1 $\sigma$ of the best fit, where $\sigma$ was determined as
described above. The quoted errors are the standard deviations in the
values. Since there are degeneracies between the different parameters,
the actual range of, for example, ages that gives an acceptable fit
can actually be larger than is suggested from these quoted errors, but
only with a certain combination of values for the other
parameters. The standard deviations given here are merely a measure of
the scatter in the parameter values for all acceptable fits.  Of the
$\sim$200,000 SCs that were fit to each OMS, around 100 (0.05\%) gave
acceptable fits (within 1$\sigma$ from the best fit) for fields C1 and
C2 and around 1000 (0.5\%) for fields O1 and O2, owing to the smaller
significance of the OMS feature in the latter two fields. To each YMS
$\sim$100,000 SCs were fit, from which around 100 (0.1\%) were within
1$\sigma$ from the best fit.

Our fit results for the OMS in the different fields are in very good
agreement with each other. The favored foreground extinction values
are close to, but slightly higher than the values from \cite{sfd} for
all fields.  All fields favor a mean stellar age of $\sim$3-6 Gyr, but
the exact age and age spread of the OMS population are not well
constrained. For example, some SCs with an age range of 5 to 12 Gyr
give acceptable fits. Given the large amount of contamination in these
fields, the existence of less numerous populations of different ages
(for example `ancient' stars) cannot be excluded. Also, for none of
the fields we can make a statistically significant distinction between
models with small or very large age spread.  An exact age and age
spread would have to be derived from the exact location and width of the
MSTO in magnitude. Because the MSTO region is heavily contaminated by
halo and thick disk stars and because of the l.o.s. extent of CMa this
information is largely lost.

The metallicity is constrained better by these fits and gives
[Fe/H]$\sim$-1.0 for the central fields and [Fe/H]$\sim$-0.6 for the
outer fields. This might hint at a metallicity gradient within CMa,
but owing to the rather large uncertainties on the metallicities of
the outer fields, significance of this is low. The best-fit values for
the central fields agree very well with the spectroscopic metallicity
measurements from N.F. Martin et al. (2007, in preparation). They
obtained FLAMES and AAOMEGA spectra of several hundred Red Clump stars
in several pointings within a few degrees of the center of CMa. For
the kinematically selected CMa members they find [Fe/H]=-0.9 with a
spread of $\sim$0.2 dex.  Our estimates for the outer fields agree
well with the metallicity found by \cite{carraro06} for the old
stellar population in a field at ($l$,$b$)=(232\degr,-6\degr).

The distance moduli found for fields C1, C2, O1, and O2 correspond
to distances of 7.4, 7.9, 6.9, and 7.2 kpc respectively, in 
good agreement with previous estimates, which range from 6.9 to 8.0
kpc \citep{bellazzini04,martin04b,martinez05,bellazzini06}. 

The l.o.s. depth of the CMa over-density (reflected in the OMS
dispersion in magnitudes) shows the same picture in all fields, with
an average best-fit value of 0.42 magnitudes.  The observed MS width
is not only caused by the spread along the l.o.s. but is also due to
the intrinsic metallicity spread of the stars, differential reddening,
and binary stars. The best-fit extinctions agree with the \cite{sfd}
values implying that practically all extincting material seems to be
located between us and the over-density, not within the
over-density. With a realistic binary fraction and a metallicity spread
of 0.2 in [Fe/H], similar to the spread measured spectroscopically by
N.F. Martin et al. (2007, in preparation), our model CMDs indicate
that the majority of the observed OMS width reflects the l.o.s. extent
of the CMa over-density. Of course, the corresponding physical
half-width, $\sigma_{los}$, in kpc depends on the absolute distance,
and is 1.5 kpc at 7.5 kpc.

Also the SC fit results for the YMS are internally consistent. Again
the extinction values are close to the \cite{sfd} values, although now
slightly lower. Since the YMS is expected to be equidistant with or
behind the OMS rather than in front of it, the extinction towards the
YMS can of course not be lower than towards the OMS in reality.

Despite the fact that for the YMS no distinct MSTO is observed,
the ages are better constrained than for the OMS, because
contamination by fore- and background stars is not a serious problem.
All fits within 1$\sigma$ of the best
fit have an upper age limit Age$_{ul}$ of 2 Gyr. Moreover, in the case
of field C2, all acceptable fits have the same age bin of 0.25-2 Gyr,
meaning that all other age bins give fits that are at least 1$\sigma$
worse than the best fit. For both fields the oldest age bin (2 to 5
Gyr) is excluded with more than 99\% confidence, arguing a posteriori
that a distinction of OMS and YMS is sensible. Also all age bins
with an Age$_{ul}$ less than 100 Myr are excluded with more than 99\%
confidence. According to our results, the youngest stars in the YMS
must be at least as young as $\sim$700 Myr and the oldest at most 2
Gyr old. Thus, from our fits the YMS seems to consist of stars with
ages between a few hundred million years and 2 Gyr. These estimates
agree with those of \cite{bellazzini04}, but are older than found by
\cite{carraro05} and \cite{moitinho06}.

The best-fit metallicity is more metal-rich than that of the OMS,
[Fe/H]$\sim$-0.3, and in that case the YMS is somewhat more distant
than the OMS, namely 9.1 and 9.5 kpc in C1 and C2, respectively, but
with large uncertainties. As will be discussed in more detail below,
due to the degeneracy between these parameters, they are not well
constrained individually.  The measured l.o.s. dispersion in field C1
is similar to that of the OMS, but a bit smaller in field C2. At their
respective best-fitting distances the recovered l.o.s. dispersions for
the YMS in C1 and C2 correspond to a $\sigma_{los}$ of $\sim$1.8 kpc
and $\sim$1.3 kpc.

Based on the best-fit parameters from Table \ref{tab:scresults} we can
reconstruct what the CMDs of our fields would look like if there were
only CMa stars present without any contamination by fore- and
background stars. Using the isochrones from \cite{girardi02}, assuming
a standard Salpeter IMF, and applying the photometric errors and
completeness of our data, gives realistic representations of our
best-fit SC models. The resulting CMDs in Figure \ref{fig:modelcmds}
show an approximate, `clean' picture of the CMa overdensity in fields
C1, C2, O1 and O2. Comparing Figure \ref{fig:modelcmds} with Figure
\ref{fig:cmacmds} nicely illustrates the amount of contamination we
have to deal with.

\subsection{Distance-metallicity degeneracy}

\begin{figure*}
\centering
\epsscale{0.8}
\plotone{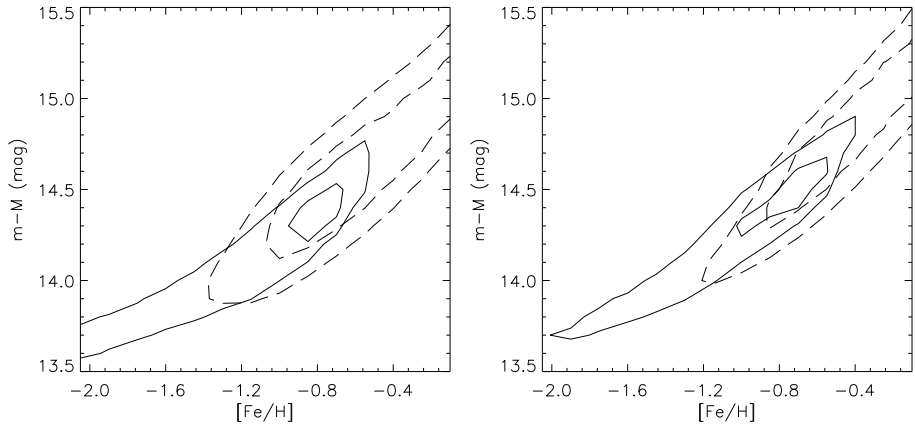}
\caption{ Contour plots of the goodness-of-fit as function of
metallicity and distance modulus of the SC fits in fields C1 (left)
and C2 (right).  The 1 and 3 $\sigma$ contours are plotted with solid
lines for the OMS and with dashed lines for the YMS. Only SC fits with
foreground extinction of $A_V$=1.1 and $A_V$=0.7 for C1 and C2
respectively were used to construct the contours.  }
\label{fig:distmetcontours}
\end{figure*}

\begin{figure*}
\epsscale{1.0}
\plotone{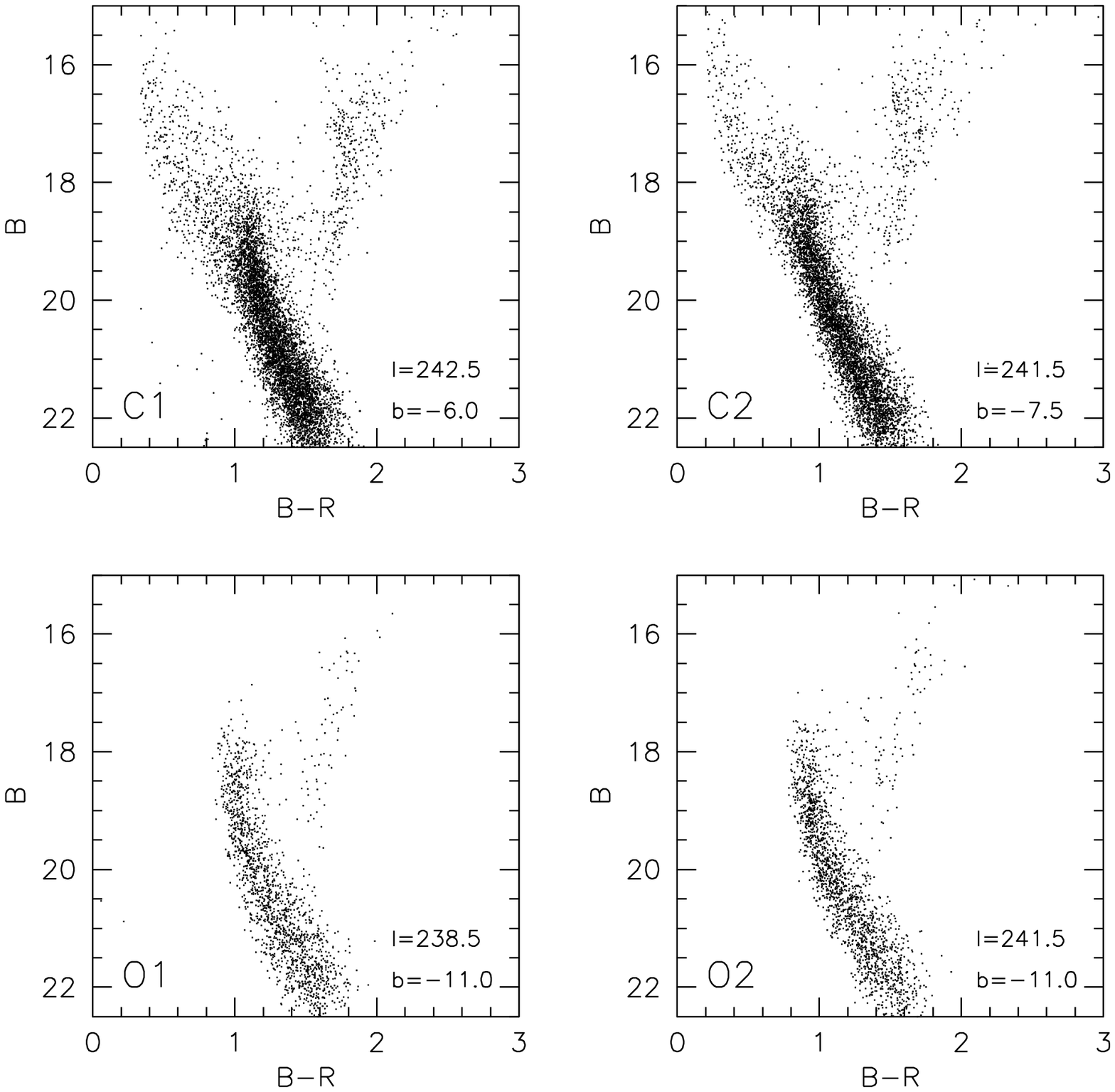}
\caption{ Model CMDs of the Canis Major over-density for the fields
  analyzed in this study. The reconstructions are based on the
  best-fit values for the OMS and YMS in Table \ref{tab:scresults} and
  the isochrones from \cite{girardi02} and include the photometric
  errors and completeness determined from the artificial star
  tests. Also the number of stars in the CMDs is scaled to reproduce
  realistic models of what the CMDs in Figure \ref{fig:cmacmds} would
  look like in the absence of fore- and background stars.}
\label{fig:modelcmds}
\end{figure*}

One of the key questions in the CMa debate is whether the young and
old stellar over-densities are physically related or not. Although the
values in Table \ref{tab:scresults} suggest at first glance that the
YMS may actually be farther away and half a dex more metal-rich than
the OMS, the degeneracy between metallicity and distance prohibits a
secure determination of this. Figure \ref{fig:distmetcontours} shows
the goodness-of-fit contours in the metallicity vs. distance modulus
plane for the YMS (dashed) and OMS (solid) in fields C1 and C2. To
remove another degeneracy, the foreground extinction was fixed to 1.1
for C1 and 0.7 for C2, but all age bins and $\sigma_{los}$ bins were
included for making these plots.  The direction of the degeneracy is
obvious, with the best-fit distance increasing for increasing
metallicity. While for these foreground extinction values the
metallicity of the OMS is relatively robustly determined at
[Fe/H]=-0.8, the metallicity constraint for the YMS is not strong.  If
the YMS is significantly more metal-rich than the OMS, it must be more
distant. However, as the contours in Figure \ref{fig:distmetcontours}
indicate, our fits are also consistent with the YMS having a similar
metallicity to the OMS, in which case it would be at the same distance
as the OMS.

\section{Combined distance solutions}
\label{sec:distfits}

While in the previous sections the YMS and OMS populations were
treated separately, we now proceed to fit the complete CMDs of fields
C1 and C2. Using the distance fitting mode of MATCH (see section
\ref{sec:cmdfitting}) we will try to confirm the distances obtained in
the previous section and look again at the possible distance offset
between the two populations.

In the distance fitting mode of the MATCH fitting, the age bins to be
used have to be specified as well as a fixed metallicity for each age
bin.  We use eight age bins with limits in log(years) of 7.5 to 8.0,
8.0 to 8.5, 8.5 to 9.0, 9.0 to 9.3, 9.3 to 9.5, 9.5 to 9.7, 9.7 to 9.9
and 9.9 to 10.1. Based on the results from the previous section we
expect to find YMS stars in some of the four youngest age bins and OMS
stars in some of the four oldest. As the YMS metallicity is not well
constrained, we consider two cases. In the first case we assign
metallicities of [Fe/H]=-1.0 to the four oldest bins, the best-fit
value from our SC fits to the OMS that is consistent with the
spectroscopic metallicities, and metallicities of [Fe/H]=-0.3 to the
four youngest bins. In the second case we assign the same metallicity,
[Fe/H]=-0.85, to all age bins. The extinction is fixed at an $A_V$ of
1.0 for C1 and 0.7 for C2 and we apply a Gaussian distance modulus
spread of 0.5 magnitudes. We use the same control field CMDs as for
the SC fits to the OMS in fields C1 and C2.

\begin{figure*}
\epsscale{1.0}
\plotone{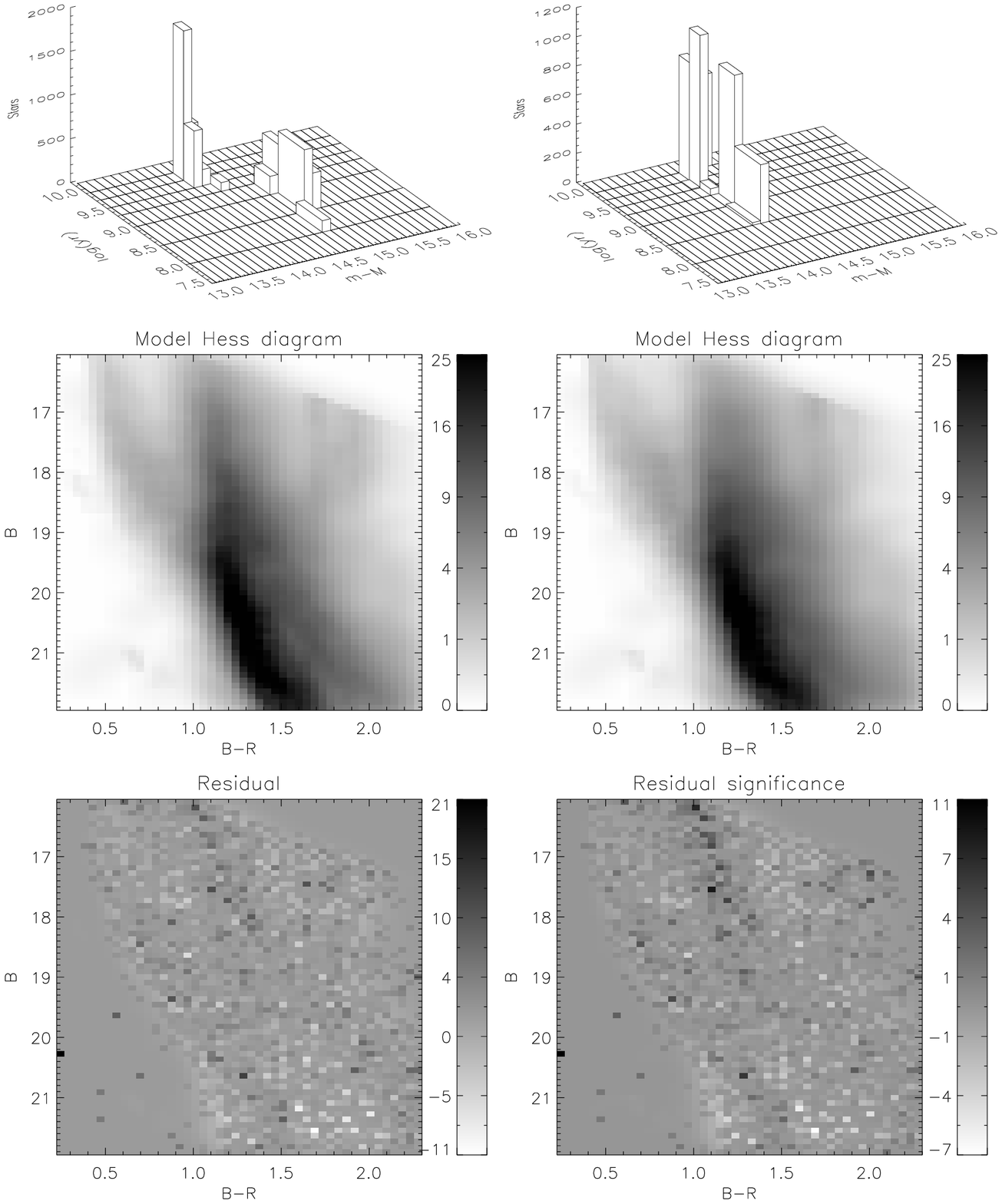}
\caption{ Distance fitting results for the complete CMD of field
  C1. {\it Left:} for the plots in this column, a metallicity of
  [Fe/H]=-1.0 was assumed for the four oldest age bins and [Fe/H]=-0.3
  for the four youngest age bins. {\it Right:} for this column a
  metallicity of [Fe/H]=-0.85 was assumed for all age bins. {\it Top:}
  best distance-age solution as 2-D histogram showing the number of
  stars in each distance-age bin needed to best reproduce the observed
  CMD; the background CMD is scaled to account for the fore- and
  background stars, meaning that the stars in these bins represent
  only the stars in the stellar over-density. {\it Middle:} the
  synthetic CMDs corresponding to the distance-age solution plotted in
  the top panels plus the background.  {\it Bottom:} significance of
  the residuals between the observed and synthetic Hess
  diagrams. Overall this shows a random scatter with only very small
  systematic deviations. In the bottom-left part (B$>$20, B-R$<$1) there
  are some pixels with high significance where stars are located in
  the control field but not in the target field. Such pixels do not
  influence the fitting since none of the probed model populations can
  reproduce these stars.
}
\label{fig:C1distfits}
\end{figure*}

\begin{figure*}
\epsscale{1.0}
\plotone{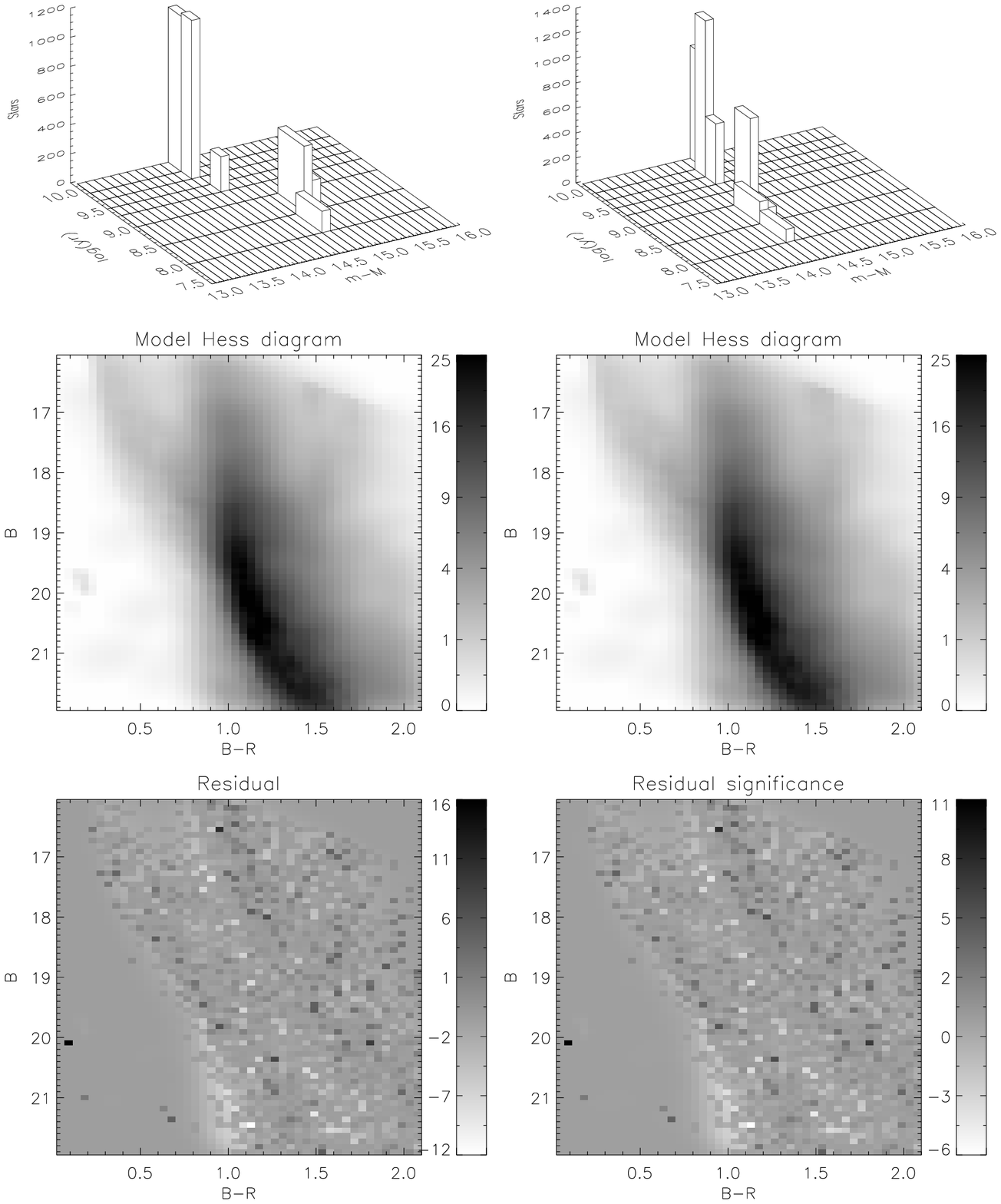}
\caption{
As Figure \ref{fig:C1distfits}, but for field C2.
}
\label{fig:C2distfits}
\end{figure*}

Figure \ref{fig:C1distfits} shows the results of the distance fits for
field C1, for the case of metal-poor OMS and metal-rich YMS (left
column), and for a uniform metallicity over the whole age range (right
column). The upper panels show the distance solutions as 2-D
histograms and the middle and lower panels show the corresponding
synthetic Hess diagrams and residual significances. In Figure
\ref{fig:C2distfits} we show the same results for field C2. The
results are in agreement with those of the previous section. Assuming
a metallicity of [Fe/H]=-1.0 for the oldest age bins and [Fe/H]=-0.3
for the youngest, the OMS stars are located at the same distance
moduli that the SC fits give, and the YMS stars are located at
distance moduli that are $\sim$0.6 magnitudes larger. Because of the
l.o.s. extent of both populations, they do partly overlap in space,
but their density peaks are clearly not equidistant in the case of
differing metallicities. As already implied by Figure
\ref{fig:distmetcontours}, when assuming a constant metallicity of
[Fe/H]=-0.85 over the whole age range, the YMS is found to be at the
same distance as the YMS.  Comparing the model Hess diagrams and
residual significance plots in Figures \ref{fig:C1distfits} and
\ref{fig:C2distfits} shows that both scenarios give practically
indistinguishable fits.

\section{Discussion}
\label{sec:discussion}

Before discussing the implications of our results for the
interpretation of the CMa stellar over-density, it is worthwhile to
review these results. 

\subsection{Overview of results}

We have investigated the stellar populations of the CMa over-density in
several fields using deep optical photometry and CMD fitting
techniques. Specifically, we explored the properties of the OMS and
YMS using SC fits in Section \ref{sec:scfits}. For the fields near the
presumed center of CMa the fits indicate a rather low metallicity of
[Fe/H]$\sim$-1.0$\pm$0.1 dex for the OMS. This agrees well with the
spectroscopic measurements from N.F. Martin et al. (2007, in
preparation), [Fe/H]$\simeq$-0.9, based on several hundred RC stars
within a few degrees from the positions of our central fields. The
distance we measure to the OMS is 7.4, 7.9, 6.9, and 7.2 kpc for
fields C1, C2, O1, and O2 respectively, which also agrees well with
previous estimates
\citep{bellazzini04,martin04b,martinez05,bellazzini06}.  We find a
$\sigma_{los}$ of $\sim$0.42 magnitudes. Since the photometric errors,
age and metallicity spread, and a realistic binary fraction are all
accounted for in the models, this directly translates into a physical
l.o.s. depth of $\sim$1.5 kpc.  This is consistent with our upper
limit for $\sigma_{los}$ for the OMS in B06. The exact stellar ages
and age distribution within the OMS population cannot be constrained
well because of the heavy contamination by fore- and background stars
in the MSTO region. But we find that the OMS population is
predominantly $\sim$3-6 Gyr old, which is typically termed an
`intermediate' age. Given the amount of contamination in the fields
this, however, does not exclude the presence of a less numerous
population of, for example, much older stars.

For the YMS the ages are better constrained and these stars have a
range in age from at least few hundred million years to at most
$\sim$2 Gyr. Based on our results we conclude with high significance
that the YMS stars cannot all be younger than 100 Myr, nor older than
2 Gyr.  Owing to a strong degeneracy between metallicity and distance
modulus (see Figure \ref{fig:distmetcontours}), a secure determination
of either of these parameters individually is precluded.  Either the
YMS has a metallicity similar to that of the OMS, in which case it
would have to be equidistant and therefore co-spatial with the OMS, or
it is more metal-rich, in which case it is located behind the OMS.
This same picture is confirmed by our distance solution fits in
Section \ref{sec:distfits}.

\subsection{Implications for the nature of CMa}

How do our findings relate to the hypotheses that have been put forward
to explain the CMa over-density? Since the possibility exists that the
OMS and YMS populations are not actually related, it should be
considered that different explanations are needed for each.  Three
theories have been suggested in the literature: (a) the CMa
over-density is (the remnant of) an dwarf galaxy being accreted onto
the MW
\citep[e.g.][]{martin04b,bellazzini04,martin05,bellazzini05,martinez05,bellazzini06};
(b) the warp and flare of the outer disk create the observed
over-density along the l.o.s. \citep{momany04,momany06}; (c) both OMS
and YMS are related to intrinsic substructure within the outer disk,
the OMS to a nearby spiral arm and the YMS to a distant spiral arm
\citep{carraro05,moitinho06}.  Below we discuss each of these options
in light of the results of this paper.

If the CMa over-density is an accreted dwarf galaxy, it would be
expected to have different properties from disk stars, for example in
kinematics and metallicity. Proper motion measurements of a sample of
YMS stars near the presumed center of CMa by \cite{dinescu05} are
consistent with this scenario \citep{monmodel}, as are kinematical
measurements of the OMS by \cite{martin04b} and \cite{martin05}.
Recent metallicity measurements of stars in the outer galactic disk by
\cite{yong06} show a clear dependence on galactocentric radius
($R_{GC}$). Our determination of the metallicity of the OMS stars of
[Fe/H]$\sim$-1.0 means that they are significantly more metal-poor
than the expectation for disk stars at $R_{GC}\sim$13 kpc,
[Fe/H]$\approx$-0.5 \citep{yong06}, thus hinting at an external
origin. The ages of the OMS stars are not well constrained, but the
majority of the population has an intermediate age of $\sim$3-6
Gyr. Many dwarf galaxies have such intermediate age populations, but
are expected to also contain a population of older
($\sim$10 Gyr) stars; but as mentioned before our results do not
exclude the presence of such stars.  The SC fits are sensitive to the
most numerous population and a less numerous population could easily
`hide' in the noise caused by the large contamination from fore- and
background stars. Unfortunately, photometry alone does not constrain
the metallicity of the YMS stars well enough and therefore we cannot say
whether or not their metallicity is consistent with them being disk
stars. Should they have similar metallicities to the OMS stars,
it would be hard to reconcile this with them being disk
stars. Furthermore, it would place them at the same distance as the
OMS stars (e.g. Figure \ref{fig:distmetcontours}), in accordance with the
picture of an accreted dwarf galaxy with an extended star
formation history. On the other hand, they might be metal-rich and
farther away than the OMS, meaning that the two populations are
probably not related at all. In this case, the metallicity and
kinematics of the OMS would still be consistent with an external
origin, but the YMS might well originate in the Galaxy.
Spectroscopic measurements of YMS stars will clearly be of key
importance to resolve this issue.

In B06 we compare the observed stellar density profile along the
l.o.s. towards Canis Major with predictions from Galactic warp/flare
models. This analysis shows that the warp of the outer disk is
unlikely to produce as narrow a peak in the stellar density at a
distance of $\sim$7.5 kpc as is observed. Rather, a more extended
density peak is expected, peaking much closer to us. The distance and
$\sigma_{los}$ determinations in this paper confirm that the
l.o.s. density profile of CMa has a narrow peak with
$\sigma_{los}\simeq$1.5 kpc centered at $\sim$7.5 kpc and are thus
inconsistent with the warp interpretation of CMa. As stated
before, the metallicity we find for the OMS is significantly lower
than expected for disk stars, also arguing against the warp/flare
hypothesis.  Although the YMS stars might be disk stars, their
distance ($\sim$9.3 kpc in this case) is also inconsistent with smooth
(locally axisymmetric) warp models. Additional substructure would have
to be invoked to explain an over-density of young stars at this
distance.

\begin{figure}
\epsscale{1.0}
\plotone{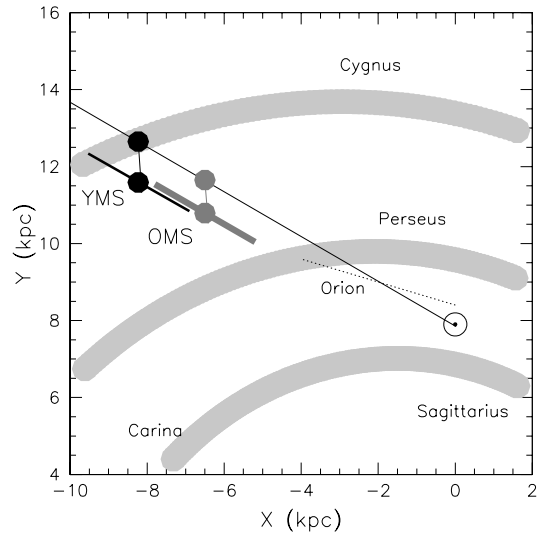}
\caption{Schematic overview of the third quadrant of the MW, seen
face-on from the North Galactic pole; the light-grey bands trace the
spiral arms according to \cite{vallee05}, based on a combination of
different spiral arm tracers. The position of the sun is indicated and
the galactic center is at (0,0) in these $X,Y$ coordinates. Drawn as a
solid line is the projection onto the Galactic plane of the
l.o.s. towards the fields used in this paper ($l$=240\degr), and the
approximate location of the local Orion arm is indicated with a dotted
line. The grey and black circle and band indicate the peak distance
and l.o.s. extent (assuming a symmetrical distribution along the
l.o.s.)  of the OMS and YMS, respectively, in the case that the YMS
has [Fe/H]=-0.3 dex. Also plotted are the peak distances projected
onto the Galactic plane, showing that for this metallicity the YMS is
equidistant with the outer spiral arm.  }
\label{fig:spirals}
\end{figure}

Large scale substructure in the disk exists in the form of spiral
arms, and especially young stellar populations are concentrated in
these. \cite{moitinho06} suggest that the young stars seen in the
direction of Canis Major and in the background of several open
clusters are part of the outer Norma-Cygnus spiral arm. In their
picture, the old stars are part of the Orion (Local) arm, which is
located in the inter-arm region between the Sagittarius and Perseus
spiral arms, meaning that the old stars are much closer to us. If the
YMS is part of a Galactic spiral arm it should be expected to be
relatively metal-rich. Assuming [Fe/H]=-0.3, our fits imply a distance
to the YMS of $\sim$9 kpc. Figure \ref{fig:spirals} shows a schematic
face-on view of the third quadrant of the Milky Way. Spiral arms
according to \cite{vallee05}, which are based on a combination of
different optical and HI spiral arm tracers, are outlined as well as
the position of the sun and the l.o.s. towards the CMa fields studied
in this paper. The black dot indicates the location of the YMS if
[Fe/H]=-0.3; it coincides with the distance to the outer spiral
arm. Our fits therefore confirm that when assuming close-to-solar
metallicities one will find that the distance to the YMS is the same
as the distance to the Norma-Cygnus spiral arm. It should be noted
that although the YMS coincides with the spiral arm in this face-on
view, it has a large offset in Z, the direction perpendicular to the
plane. If this distance is correct, these YMS stars are located more
than 1 kpc out of the plane of the disk. All ``classic'' tracers of
the spiral arm are less than 0.5 kpc out of the Galactic midplane
($b$=0), with the exception of only one detected molecular cloud. The
proper motion measurements by \cite{dinescu05} show that the YMS stars
are moving even farther away from the plane. Some additional
explanation for the extreme location and kinematics of these stars is
needed to reconcile them with a Galactic origin.  If we assume that
the metallicity of the YMS is much lower than solar and similar to
that of the OMS, the best-fit distance of the YMS shifts to the same
distance as the OMS, in which case it is located in an inter-arm
region. Again, spectroscopic measurement of the metallicity of the YMS
is crucial to solve this problem.

The location of the OMS is indicated in Figure \ref{fig:spirals} with
the dark grey dot; it lies in the inter-arm region between the Perseus
and Cygnus spiral arms. Also drawn in Figure \ref{fig:spirals} is the
approximate position of the Orion arm, as outlined by
\cite{moitinho06} in their Figure 2, based on the positions of open
star clusters. Clearly, the Orion arm crosses the l.o.s. towards our
fields at a much smaller distance of approximately 2 kpc, or a
distance modulus of m-M=11.5. Such a distance is far smaller than what
previous studies of CMa have indicated and excluded at high confidence
by our results (see e.g. Table \ref{tab:scresults} and Figure
\ref{fig:distmetcontours}). Furthermore, the metallicity we find for
the OMS in fields C1 and C2, [Fe/H]$\sim$-1.0, is much lower than is
found for the Galactic thin and thick disk
\citep[e.g.][]{yong06,bensby04}.

\section{Summary and Conclusions}
\label{sec:conclusions}

Applying CMD-fitting techniques to a small subset of fields from our
survey of the CMa region (B06) we have determined the distance,
l.o.s. extent and metallicity of the old and young stars in the CMa
over-density. For the ``old'' stars (OMS) the recovered distance,
$\sim$7.5 kpc, $\sigma_{los}$, $\sim$1.5 kpc, metallicity,
[Fe/H]$\sim$-1.0, and mean ages, $\sim$3-6 Gyr, are in good agreement
with results from the literature.  Also, we have constrained the
metallicity, distance, l.o.s. extent and ages of the younger
population of stars seen in the same direction. These stars have ages
of at least a few hundred million years and at most 2 Gyr. The
degeneracy between distance and metallicity prevents us from drawing a
firm conclusion about the co-spatiality of the old and young stars. A
metal-rich ([Fe/H]=$\sim$-0.3) population at $\sim$9.3 kpc and a
metal-poor ([Fe/H]=$\sim$-0.8) population at 7.5 kpc are both
consistent with the data. Spectroscopic metallicities for the young
stars are necessary to distinguish between these two possibilities.

A comparison of the distance of the OMS with Galactic warp/flare
models argues against the interpretation that the OMS is the result of
the warp and flare of the Galactic disk (B06). Our results are
inconsistent with the OMS stars being located in the local Orion arm,
since the implied distance of $\sim$2 kpc is ruled out with high
significance. The metallicity we find seems too low for disk stars,
hinting at an external origin.  Because of the degeneracy mentioned
above we cannot be certain about the nature of the YMS. If it is
metal-rich, it coincides in distance, but not in height above the
disk, with the Norma-Cygnus spiral arm. Perhaps a scenario where a
combination of spiral arm structure and the warp conspire to produce
young stars so far out of the disk is feasible.  On the other hand, if
the YMS has a similar metallicity to the OMS, it is co-spatial with
the OMS and located in an inter-arm region.  In this case, the
accreted dwarf galaxy scenario would be the more likely explanation.

\acknowledgments

We would like to thank N.F. Martin for fruitful discussions and for
providing preliminary spectroscopic metallicities. 
We also thank the anonymous referee for many helpful suggestions and
comments that improved the overall quality of this paper.
JTAdJ and DJB acknowledge support from DFG Priority Program 1177.
DMD recognizes the support of the Spanish Ministry of Education and
Science (Ramon y Cajal contract and research project AYA
2001-3939-C03-01).


\begin{thebibliography}{}
\bibitem[Aparicio, Gallart \& Bertelli(1997)]{aparicio97} Aparicio,
  A., Gallart, C. \& Bertelli, G. 1997, \aj, 114, 680
\bibitem[Bellazzini et al.(2004)]{bellazzini04} Bellazzini, M., Ibata,
  R., Monaco, L., Martin, N., Irwin, M. J., Lewis, G. F., 2004,
  \mnras, 354, 1263
\bibitem[Bellazzini et al.(2005)]{bellazzini05} Bellazzini, M., Ibata,
  R. A., Monaco, L., Martin, N., Irwin, M. J., et al. 2005, \mnras,
  354, 1263
\bibitem[Bellazzini et al.(2006)]{bellazzini06} Bellazzini, M., Ibata,
  R. A., Martin, N., Lewis, G. F., Conn, B., Irwin, M. J., 2006,
  \mnras, 366, 865
\bibitem[Bensby et al.(2004)]{bensby04} Bensby, T., Feltzing, S. \&
  Lundstr\"om, I., 2004, \aap, 421, 969
\bibitem[Bertin \& Arnouts (1996)]{sextractor} Bertin, E. \& Arnouts,
  S., 1996, \aaps, 117, 393
\bibitem[Bonifacio et al.(2000)]{bonifacio00} Bonifacio, P., Monai,
  S., Beers, T. C., 2000, \aj, 120, 2065
\bibitem[Butler et al.(2006)]{butler06} Butler, D. J., Martinez-Delgado,
  D., Rix, H-W., Pe\~narrubia, J., De Jong, J. T. A., 2006, subm.,
  astro-ph/0609316, B06
\bibitem[Carraro et al.(2005)]{carraro05} Carraro, G.,
  V\'azquez, R.~A., Moitinho, A. \& Baume, G., 2005,
  \apjl, 630, L153
\bibitem[Carraro et al.(2006)]{carraro06} Carraro, G., Moitinho, A.,
  Zoccali, M., V\'azquez, R. A. \& Baume, G., 2006, \aj, in press,
  astro-ph/0610617
\bibitem[Dinescu et al.(2005)]{dinescu05} Dinescu, D. I.,
	Mart\'{\i}nez-Delgado, D., Girardi, T. M., Pe{\~n}arrubia, J.,
	Rix, H.-W., Butler, D. J., van Altena, W. F., 2006, \apj, 631, L49
\bibitem[Dolphin(1997)]{dolphin97} Dolphin, A. E. 1997, \na, 2, 397
\bibitem[Dolphin(2002)]{match} Dolphin, A. E. 2002, \mnras, 332, 91
\bibitem[Duquennoy \& Mayor(1991)]{duquennoy91} Duquennoy, A. \&
  Mayor, M., 1991, \aap, 248, 485
\bibitem[Gallart et al.(1996)]{gallart96} Gallart, C., Aparicio, A.,
  Bertelli, G., Chiosi, C. 1996, \aj, 112, 1950
\bibitem[Girardi et al.(2002)]{girardi02} Girardi, L., Bertelli, G.,
  Bressan, A., Chiosi, C., Groenewegen, M. A. T., Marigo, P.,
  Salasnich, B. \& Weiss, A., 2002, \aap, 391, 195
\bibitem[Harris \& Zaritsky(2001)]{harris01} Harris, J. \& Zaritsky,
  D. 2001, \apjs, 136, 25
\bibitem[Hernandez, Gilmore \& Valls-Gabaud(2000)]{hernandez00}
  Hernandez, X., Gilmore, G. \& Valls-Gabaud, D. 2000, \mnras, 317, 831
\bibitem[Holtzman et al.(1999)]{holtzman99} Holtzman, J. A. et al.,
  1999, \aj, 118, 2262
\bibitem[Kroupa et al.(1993)]{kroupa93} Kroupa, P., Tout, C. A. \&
  Gilmore, G., 1993, \mnras, 262, 545
\bibitem[Martin et al.(2004a)]{martin04a} Martin, N. F., Ibata, R. A.,
  Bellazzini, M., Irwin, M. J., Lewis, G. F., Dehnen, W., 2004a,
  \mnras, 348, 12
\bibitem[Martin et al.(2004b)]{martin04b} Martin, N. F., Ibata, R. A.,
  Conn, B. C., Lewis, G. F., Bellazzini, M., Irwin, M. J.,
  McConnachie, A. W., 2004b, \mnras, 355, L33
\bibitem[Martin et al.(2005)]{martin05} Martin, N. F., Ibata, R. A.,
  Conn, B. C., Lewis, G. F., Bellazzini, M., Irwin, M. J., 2005,
  \mnras, 362, 906
\bibitem[Martinez-Delgado et al.(2005)]{martinez05}
  Mart\'inez-Delgado, D., Butler, D. J., Rix, H-W., Franco, Y. I.,
  Pe\~narrubia, J., 2005, \apj, 633, 205
\bibitem[Moitinho et al.(2006)]{moitinho06} Moitinho, A., V\'azquez,
  R. A., Carraro, G., Baume, G., Giorgi, E. E., Lyra, W., 2006,
  \mnras, 368, L77
\bibitem[Momany et al.(2004)]{momany04} Momany, Y., Zaggia, R. S.,
  Bonifacio, P., Piotto, G., De Angeli, F., Bedin, L. R., Carraro, G.,
  2004, \aap, 421, L29
\bibitem[Momany et al.(2006)]{momany06} Momany, Y., Zaggia, R. S.,
  Gilmore, G., Piotto, G., Carraro, G., Bedin, L. R., De Angeli, F.,
  2006, \aap, 451, 515
\bibitem[Newberg et al.(2002)]{newberg02} Newberg, H. J., Yanny, B.,
  Rockosi, C., Grebel, E. K., Rix, H-W., et al., 2002, \apj, 569, 245
\bibitem[Olsen(1999)]{olsen99} Olsen, K. A. G. 1999, \aj, 117, 2244
\bibitem[Pe\~narrubia et al.(2005)]{monmodel} Pe\~narrubia, J.,
  Mart\'inez-Delgado, D., Rix, H. W., G\'omez-Flechoso, M. A., Munn,
  J., 2005, \apj, 626, 128
\bibitem[Rocha-Pinto et al.(2006)]{rocha06} Rocha-Pinto, H. J.,
	Majewski, S. R., Skrutskie, M. F., Patterson, R. J.,
	Nakanishi, H., Mu{\~n}oz, R. R., Sofue, Y., 2006, \apj, 640, L147
\bibitem[Schirmer et al.(2003)]{schirmer03} Schirmer, M., Erben, T.,
  Schneider, P., Pietrzynski, G., Gieren, W., 2003, \aap, 407, 869
\bibitem[Schlegel, Finkbeiner \& Davis(1998)]{sfd} Schlegel, D.,
  Finkbeiner, D. \& Davis, M. 1998, \apj, 500, 525
\bibitem[Tolstoy \& Saha(1996)]{tolstoy96} Tolstoy, E. \& Saha,
  A. 1996, \apj, 462, 672
\bibitem[Vallee(2005)]{vallee05} Vallee, J. P., 2005, \aj, 130, 569
\bibitem[Yanny et al.(2003)]{yanny03} Yanny, B., Newberg, H. J.,
  Grebel, E. K., Kent, S., Odenkirchen, M., 2003, \apj, 588, 824
\bibitem[Yong et al.(2006)]{yong06} Yong, D., Carney, B. W., Teixera
  de Almeida, M. L., Pohl, B. L., 2006, \aj, 131, 2256
\end{thebibliography}
\end{document}